\documentclass[10pt,conference]{IEEEtran} 

\usepackage{amsfonts}
\usepackage{amssymb}
\usepackage{cite}
\usepackage{url}
\usepackage[]{algorithm2e}
\usepackage{balance}
\usepackage{enumerate}
\usepackage{latexsym}
\usepackage{rotating}
\usepackage{float}
\usepackage{graphicx}
\usepackage{paralist}
\usepackage{multirow}
\usepackage{multicol}
\usepackage{pifont}
\usepackage{amsmath}
\usepackage[T1]{fontenc}
\usepackage[usenames]{color}
\usepackage{fancybox}
\usepackage{arydshln} %to use \hdashline
\makeatletter

\newcommand{\MYCOMMENT}[1]{}

\usepackage{xspace}

\newcommand{\hypobox}[1]{\begin{center}%	
	\noindent\thicklines\setlength{\fboxsep}{8pt}%	
	\cornersize{0.2}\Ovalbox{\begin{minipage}{3.0in}%	
	\textit{#1}\end{minipage}} \end{center}}

\newcommand{\etal}{\textit{et al.}\xspace}

\begin{document}

\title{Comprehension of Ads-supported and\\Paid Android Applications: Are They Different?}

\author{
    \IEEEauthorblockN{Rub\'{e}n Saborido, Foutse Khomh, Yann-Ga\"{e}l Gu\'{e}h\'{e}neuc, Giuliano Antoniol}
     \IEEEauthorblockA{Polytechnique Montr\'{e}al, Canada}
}

% make the title area
\maketitle

\begin{abstract}
The Android market is a place where developers offer paid and--or free apps to users. Free apps can follow the freemium or the ads-business model. While the former offers less features and the user is charged for unlocking additional features, the latter includes ads to allow developers to get a revenue. Free apps are interesting to users because they can try them immediately without incurring a monetary cost. However, free apps often have limited features and--or contain ads when compared to their paid counterparts. Thus, users may eventually need to pay to get additional features and--or remove ads. While paid apps have clear market values, their ads-supported versions are not entirely free because ads have an impact on performance. The hidden costs of ads, and the recent possibility to form family groups in Google Play to share purchased apps, make it difficult for developers and users to balance between visible and hidden costs of paid and ads-supported apps. 

In this paper, first, we perform an exploratory study about ads-supported and paid apps to understand their differences in terms of implementation and development process. 
We analyze 40 Android apps and we observe that (i) ads-supported apps are preferred by users although paid apps have a better rating, (ii) developers do not usually offer a paid app without a corresponding free version, (iii) ads-supported apps usually have more releases and are released more often than their corresponding paid versions, (iv) there is no a clear strategy about the way developers set prices of paid apps, (v) paid apps do not usually include more functionalities than their corresponding ads-supported versions, (vi) developers do not always remove ad networks in paid versions of their ads-supported apps, and (vii) paid apps require less permissions than ads-supported apps. Second, we carry out an experimental study to compare the performance of ads-supported and paid apps and we propose four equations to estimate the cost of ads-supported apps. We obtain that (i) ads-supported apps use more resources than their corresponding paid versions with statistically significant differences and (ii) paid apps could be considered a most cost-effective choice for users because their cost can be amortized in a short period of time, depending on their usage.
\end{abstract}

\begin{IEEEkeywords}
Android; Performance metrics; Advertisements
\end{IEEEkeywords}

\newcommand{\rqmain}{Are ads-supported and paid Android apps different?}

\newcommand{\rqmaina}{How is the ads-supported business model used?}

\newcommand{\rqmainb}{How is the ads-supported business model impacting performance and price?}

%Questions for first main RQ.
\newcommand{\rqone}{Do developers prefer ads-supported or paid apps?}

\newcommand{\rqtwo}{Do users prefer ads-supported or paid apps?}

\newcommand{\rqthree}{How often are ads-supported apps released?}

\newcommand{\rqfour}{How does the price of paid apps evolve?}

\newcommand{\rqfive}{Do ads-supported apps have reduced features?}

\newcommand{\rqsix}{What are the ad networks used?}

\newcommand{\rqseven}{What is the impact of ads on permissions?}

%Questions for second main RQ.
\newcommand{\rqeight}{Is the impact of ads on performance significant?}

\newcommand{\rqnine}{Is the impact of ads on price significant?}

\section{Introduction}
\label{sec:introduction}
The Android market Google Play is the place where Android developers and users meet to offer and to find apps. In this market, developers may wish to maximize the number of downloads and ratings of their apps while users may be more interested in finding the best and cheapest apps that fulfill their needs. In general, developers want to maximize their profits while users  want to reduce their costs.

There exist different business models for Android apps\footnote{https://developer.android.com/distribute/monetize/index.html}. One model is the pay to download in which users pay the app before downloading. The opposite model is the free app-only model, in which developers do not seek any monetary reward but, most likely, coverage for their apps. Between these two extreme models stand the freemium model, in which developers offer apps for free but ask money for unlocking more features, and the ads-supported model, in which developers offer apps for free but use ads to generate revenues.

On the one hand, both developers and users are interested by free apps: developers to showcase their apps and users to test out these apps for free. On the other hand, developers may offer, in addition to their free versions, paid apps and include in the free versions ads. These ads-supported apps offer less or similar features than their corresponding paid versions and they use ad networks to display ads that provide revenue to developers. While paid apps have clear market values (their prices), ads-supported versions are not entirely free because ads in apps have an impact on app ratings and users' privacy \cite{book_longitudinal_2013} but also on performance metrics as CPU, memory, network usage, and power consumption \cite{wei_profiledroid:_2012,gui_truth_2015}. 

Yet, users are sometimes reluctant to pay for apps when ads-supported versions of the same or similar apps exist for free\footnote{https://goo.gl/31IXuZ}. For this reason, and to increase the numbers of purchases of paid apps, Google launched in August 2016 the concept of ``Family Group"\footnote{https://support.google.com/googleplay/answer/6286986}. When  users set up a family group on Google Play, the family manager can invite up to five people to the group and they can share purchased apps. Thus, the prices of the purchased apps is divided by the numbers of family members. 

To the best of our knowledge, there are currently no recommendations for developers regarding the business model to choose for their apps and the impact of this choice. In this paper we want to understand the usage of the ads-supported model by developers and users and, consequently, we ask the general question: \emph{``\rqmain''}. We answer this question in two steps, by asking: \emph{``\rqmaina''} and \emph{``\rqmainb''}

We answer the first question through an exploratory study about the ads-business model comparing ads-supported and paid apps to understand their differences and development process. We analyze 40 Android apps, 20 ads-supported apps downloaded from Google Play and their corresponding paid versions bought in the same market. We analyze the frequency of releasing of ads-supported and paid apps and the evolution of the prices of paid apps across releases. In addition, we collect and process information about all the developed Android apps offered in Google Play by developers of the selected apps to know the proportion of paid apps with respect to free ones in the marketplace. We also compare both ads-supported and paid apps in terms of required permissions, used ad networks, and offered features.

We answer the second research question by carrying out an experimental study to compare the performance of ads-supported and paid apps and we propose four equations to estimate the cost of ads-supported apps. We want to make explicit the hidden costs of ads when considering the possibility to form family groups. Thus, we want to provide some advices to developers, to seize and act on the balance between visible and hidden costs of paid and ads-supported apps. We collect performance metrics running each app in a real phone (repeating 30 times each measure to allow for statistical tests and effect size). Using these data, we determine the cost of ads-supported apps due to ads depending on the network usage, the battery drained, and the time in which a data plan is over, to estimate the time in which an ads-supported app overtakes its paid version.

From our exploratory study, we observe that (i) ads-supported apps are preferred by users although paid apps have a better rating, (ii) developers do not usually offer a paid app without a corresponding free version, (iii) ads-supported apps usually have more releases and are released more often than their corresponding paid versions, (iv) there is no a clear strategy about the way developers set prices of paid apps, (v) paid apps do not usually include more features than their corresponding ads-supported versions, (vi) developers do not always remove ad networks in paid versions of their ads-supported apps, and (vii) usually paid apps require less permissions than ads-supported apps.

From our experimental study, we confirm that (i) ads-supported apps use more resources than their corresponding paid versions and we conclude that differences are statistically significant and (ii) we show that, depending on the usage, paid apps could be considered as a better choice for users because its cost could be amortized in a short period of time.

The remainder of the paper is organized as follows. \mbox{Section \ref{sec:case-study}} presents the data collection process used in our exploratory research and experimental study. Section \mbox{\ref{sec:exploratory-research}} describes the experiments carried out and the observations found by our exploratory research. \mbox{Section \ref{sec:experiments}} presents and discusses the experimental study about performance and  cost of ads-supported apps. \mbox{Section \ref{sec:threat}} discusses threats to the validity of our study. Finally, \mbox{Section \ref{sec:related-work}} reports related works and \mbox{Section \ref{sec:conclusion}} concludes, summarizing and discussing our findings and providing some advices to developers.

\section{Study Data Collection}
\label{sec:case-study}
We want to analyze ads-supported and paid versions of Android apps to understand the usage of the ads-business model by developers and users. In addition, we want to study the balance between the costs of ads in free apps and the costs of paid apps, while considering their features and sharing among family groups. Finally, we want to provide developers with some advices about the impact of the business model on both users and developers.

\subsection{Selection of Subject Apps}

The context of our study is the official Android marketplace, Google Play, and the subjects are ads-supported and paid Android apps available in this marketplace. For each category in Google Play, we randomly selected eight paid apps with an ad-supported version available. Over the resulting 128 apps we selected 65 which worked on our phone and contained visible ads. Among these 65, we randomly selected 20 from different developers. Our selection process is akin to a stratified random sampling of paid apps with two strata: ad-supported version and working on our phone. We bought the paid apps and we downloaded their corresponding ads-supported versions.

Selected apps belong to 11 different categories and the number of downloads for ads-supported and paid apps was in the range $[500, 5,000,000]$ and $[50, 1,000,000]$, respectively. For all of this we consider that we have a reduced but representative sample of Android apps to carry out an exploratory research and an experimental study.

\mbox{Table \ref{table:applications}} shows the apps selected for our study. The second column shows the identifier associated with each app. The third and fourth columns contain information related to the category to which the app belongs and the developer who develops the app, respectively. Next column shows information about the versions and the number of downloads in the marketplace for the selected apps. All of this information was obtained from Google Play in May 2016, when apps were downloaded (ads-supported apps) and bought (paid apps).

\begin{table*}[!htb]
\centering
\begin{tiny}
\caption{Apps selected for the study.}
\label{table:applications}
\begin{tabular}{l l l l | l l l | l l l }%l}
\hline
\multicolumn{4}{l}{}                                                               & \multicolumn{3}{|c}{Ads-supported version}                      & \multicolumn{3}{|c}{Paid version}                         \\
\hline
App                             & Id  & Category           & Developer               & Version   & Rating & Downloads                & Version   & Rating & Downloads          \\ % & Price \\
\hline
AccuWeather Platinum            & A01 & Weather            & Accuweather.com       & 3.4.2.12  & 4.3    & 50.000.000 - 100.000.000 & 4.0.1     & 4.2    & 100.000 - 500.000   \\%& 3.90  \\
Pocket Salsa                    & A02 & lifestyle          & Modernistik           & 2.14      & 4.3    & 100.000 - 500.000        & 2.14      & 4.8    & 10.000 - 50.000     \\%& 2.49  \\
DroidEdit Pro (code editor)     & A03 & Productivity       & Andr\'{e} Restivo         & 1.23.1    & 4.1    & 1.000.000 - 5.000.000    & 1.23.1    & 4.6    & 50.000 - 100.000    \\%& 2.96  \\
How to Tie a Tie Pro            & A04 & Books \& Reference & Artfonica             & 2.6.2     & 4.6    & 5.000.000 - 10.000.000   & 2.7       & 4.6    & 10.000 - 50.000    \\% & 3.98  \\
Sleep Time+ Smart Alarm Clock   & A05 & Health \& fitness  & Azumio Inc.           & 1.36.1026 & 4.1    & 1.000.000 - 5.000.000    & 1.36.1026 & 4.2    & 1.000 - 5.000       \\%& 2.47  \\
Perfect Body Building Plan Pro  & A06 & Health \& fitness  & Bodybuilding-Apps.com & 1.2.0     & 4.2    & 100.000 - 500.000        & 1.1.6     & 4.6    & 1.000 - 5.000       \\%& 4.08  \\
CountDownr                      & A07 & Tools              & SpeedyMarks           & 1.1.0     & 2.6    & 5.000 - 10.000           & 1.1.0     & 3.5    & 100 - 500           \\%& 1.49  \\
Dictionary.com Premium          & A08 & Books \& Reference & Dictionary.com, LLC   & 5.2.2     & 4.6    & 10.000.000 - 50.000.000  & 5.2.2     & 4.7    & 100.000 - 500.000  \\% & 4.77  \\
Calculator Plus                 & A09 & Productivity       & Digitalchemy, LLC     & 4.9.2     & 4.5    & 10.000.000 - 50.000.000  & 4.9.4     & 4.8    & 50.000 - 100.000    \\%& 2.60  \\
Moon+ Reader Pro                & A10 & Books \& Reference & Moon+                 & 3.5.3     & 4.4    & 10.000.000 - 50.000.000  & 3.5.3     & 4.7    & 500.000 - 1.000.000 \\%& 6.66  \\
PRO PDF Reader                  & A11 & Books \& Reference & Ivan Ivanenko         & 4.4.1     & 4.0    & 10.000.000 - 50.000.000  & 4.4.4     & 4.4    & 10.000 - 50.000    \\% & 3.49  \\
Popster                         & A12 & Finance            & ITIOX                 & 2.1       & 3.7    & 10.000 - 50.000          & 2.1       & 4.3    & 500 - 1.000        \\% & 1.16  \\
Video Player HD Pro             & A13 & Media \& Video     & wowmusic              & 1.0.8     & 4.4    & 1.000.000 - 5.000.000    & 1.0.2     & 4.5    & 1.000 - 5.000      \\% & 2.32  \\
Time Trial Stopwatch            & A14 & Sports             & Liuto Apps            & 1.2.2     & 3.3    & 500 - 1.000              & 1.2.2     & 3.5    & 50 - 100            \\%& 1.23  \\
Mighty Grocery Shopping List    & A15 & Shopping           & Mighty Pocket         & 4.0.15    & 4.2    & 100.000 - 500.000        & 4.0.15    & 4.5    & 50.000 - 100.000    \\%& 3.99  \\
Hair Color Studio Premium       & A16 & lifestyle          & ModiFace              & 1.4       & 3.3    & 5.000.000 - 10.000.000   & 1.4       & 3.9    & 10.000 - 50.000    \\% & 1.99  \\
Date / Calendar Converter Full  & A17 & Tools              & NoM                   & 1.7       & 4.5    & 5.000 - 10.000           & 1.4       & 5.0    & 100 - 500           \\%& 1.30  \\
OBDII Trouble Codes             & A18 & Transportation     & NoM                   & 1.14      & 3.9    & 100.000 - 500.000        & 1.13      & 4.5    & 10.000 - 50.000     \\%& 5.23  \\
Maths Formulas                  & A19 & Books \& Reference & NSC Co.               & 9.1.1     & 4.4    & 1.000.000 - 5.000.000    & 9.1       & 4.6    & 1.000 - 5.000      \\% & 2.49  \\
Relax Melodies P: Sleep \& Yoga & A20 & Health \& Fitness  & Ipnos Software        & 3.2       & 4.4    & 5.000.000 - 10.000.000   & 3.3       & 4.8    & 100.000 - 500.000   \\%& 2.99 \\
\hline
\end{tabular}
\end{tiny}
\end{table*}

\subsection{Analysis and Monitoring of Subject Apps}

We collect information about releases from Google Play and AppBrain. The latter is a public source for information about Android apps where we can find data on all apps on Google Play, resources for developers, and statistics about the Android ecosystem. We obtain information about permissions using the Android tools \texttt{aapt} and \texttt{dumpsys}. To analyze ad networks in Android apps we use the Android app AppBrain Ad Detector, which is available for free in Google Play.

To compare ads-supported and paid apps in terms of performance metrics, for each app, we create a Robotium scenario to run an exercise the apps. Robotium\footnote{https://github.com/RobotiumTech/robotium} is an open-source test automation framework for Android development that makes it easy to write powerful and robust automatic black-box UI tests for Android apps. We use the commercial easy-to-use tool Robotium Recorder\footnote{http://robotium.com/products/robotium-recorder} to create Robotium test cases based on component identifiers and not on absolute or relative coordinates.

The scenarios typically start the apps, skip the initial tutorials if they exist, and wait for 100 seconds in the main activity which contains visible ads in the free apps. The same activity is used for the paid app to make a fair comparison. Because the default refresh rate value of ad networks use to be 60 seconds, setting the length of the scenario to 100 seconds guarantees that ads are loaded at least once. In our experiments, the scenarios are run 30 times while performance metrics are collected. 

We use a LG Nexus 4 Android phone, equipped with the Android Lollipop operating system (version 5.1.1, build number LMY47V), which is connected to an external power supplier to get power consumption measurements using a hardware based approach. CPU, memory, and network usages are collected using well known methods. All of this is commented in detail in \mbox{Section \ref{sec:experiments}}.

\section{Exploratory Study}
\label{sec:exploratory-research}
We refine our first high-level research question, \textit{\rqmaina}, into seven detailed research questions (\textit{RQs}). In this section, we introduce and discuss the details of the experiments that we carried out to address each of the \textit{RQs}. For each research question we describe our motivations, the data-collection approach, and discussions of our observations.

\subsection*{RQ 1: \rqone}

\subsubsection*{Motivation}

Developers can offer in Google Play independent free apps (including or not ads), independent paid apps, or both free and paid versions of the same app. We analyze the number and type of apps offered in the marketplace by the developers of the apps shown in Table \ref{table:applications} to understand what type of apps developers usually prefer. 

\subsubsection*{Approach}

We extract from Google Play information about the existing apps in the marketplace offered by developers. Then, we count the numbers of independent free apps, of independent paid apps, and of free apps with a corresponding paid version.

\subsubsection*{Results}

In total, developers offer 331 apps in Google Play. Out of these 331 apps, 116 free apps have the corresponding paid versions, 88 are without paid apps, and 11 are without free version. Thus, we conclude that (i) 70.09\% of the free apps have corresponding paid versions, (ii) 26.59\% of apps do not have corresponding paid versions, and (iii) only 3.32\% of paid apps do not have corresponding free versions.

\hypobox{Developers prefer offering both free and paid version of their apps in Google Play.} 

\subsection*{RQ 2: \rqtwo}

\subsubsection*{Motivation}

When considering ads-supported and paid apps, users can decide to install the free app that contains ads or buy the paid app without ads. We compare ads-supported and paid apps in terms of numbers of downloads and users' ratings, which can be considered as measures of success \cite{harman_app_2012,gomes_empirical_2016}. This information is useful for developers to know the point of view of users about ads-supported apps and their paid versions.

\subsubsection*{Approach}

In Google Play, the numbers of downloads and users' ratings are shown for each app in the marketplace. A rating is a number between one and five calculated as the weighted average of all users' ratings. The number of downloads is given as a range specifying how many times an app was downloaded and installed. For each ads-supported and paid app we compare these two metrics.

\subsubsection*{Results}

Table \ref{table:applications} shows that ads-supported apps are always downloaded more than their paid counterparts. Considering the central values of the ranges defining the numbers of downloads we conclude that ads-supported apps are downloaded, in average, 113 times more than their corresponding paid versions. Regarding ratings, for 19 apps (90\%), paid versions have better ratings than their corresponding ads-supported versions. For only two apps (10\%), free versions have ratings greater than or equal to their paid versions. In average, ratings for paid apps is 4.43 while for ads-supported apps is 4.09.

\hypobox{Users prefer ads-supported apps although they rate paid apps better.}

\subsection*{RQ 3: \rqthree}

\subsubsection*{Motivation}

In Google Play, developers can release their apps as often as they consider it necessary. We analyze the frequency of releasing of ads-supported and paid apps to understand how developers release free and paid apps.

\subsubsection*{Approach}

We crawl the AppBrain website\footnote{http://www.appbrain.com} to extract information about the previous releases of the studied apps. The collected information is stored in a CSV file that we use to compare the frequency of releasing of both ads-supported and their corresponding paid apps.

\subsubsection*{Results}

\begin{figure}[!htb] 
    \centering
    \includegraphics[width=0.99	\linewidth]{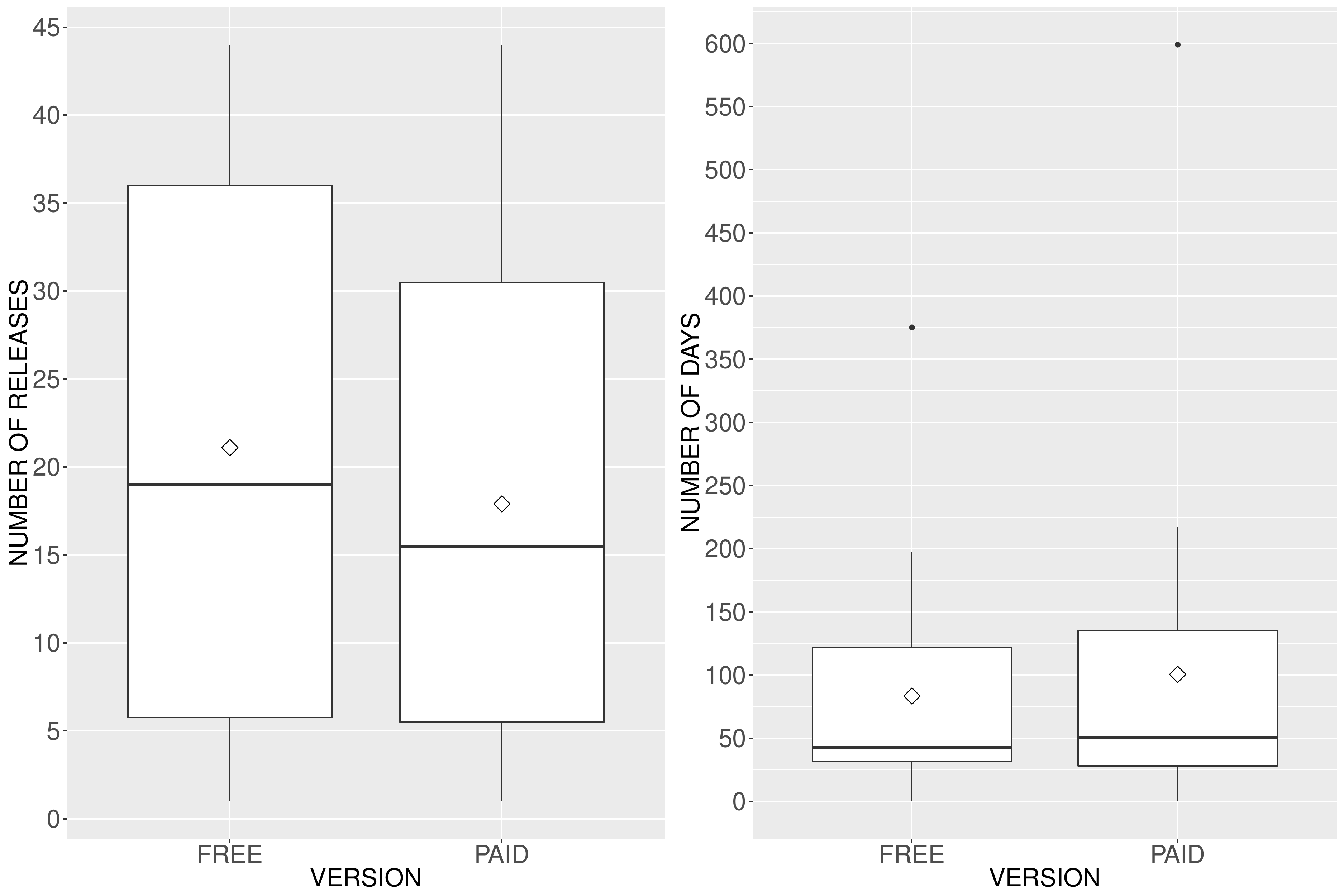} 
    \caption{Release frequencies for ads-supported and paid apps.} 
    \label{fig:ReleasesAndDays} 
\end{figure}

Fig.\ \ref{fig:ReleasesAndDays} shows the distribution of the frequencies of releasing of ads-supported and paid versions of the subject apps, where the symbol $\lozenge$ shows the average value. In average, ads-supported apps have 21.10 releases while paid apps have 17.90 releases. We also study the number of days between consecutive releases for ads-supported and paid apps. Ads-supported apps are released, in average, every 84 days (median 43 days) while paid apps are released less frequently: every 101 days in average (median 51 days). In addition, we observe that developers usually release ads-supported apps before their corresponding paid versions.

\begin{figure}[!htb] %!htb 
    \centering
    \includegraphics[width=0.8\linewidth]{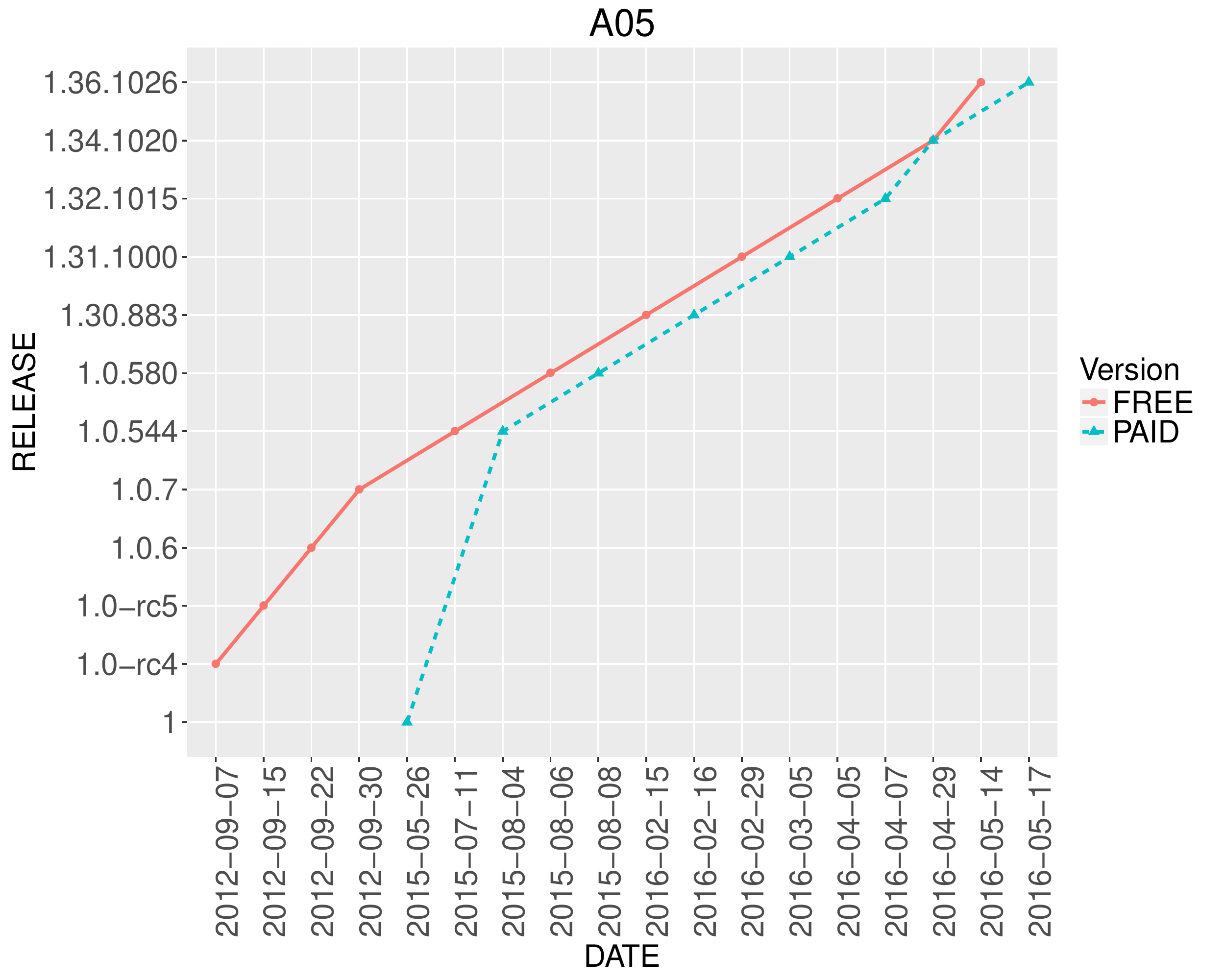}
    \includegraphics[width=0.8\linewidth]{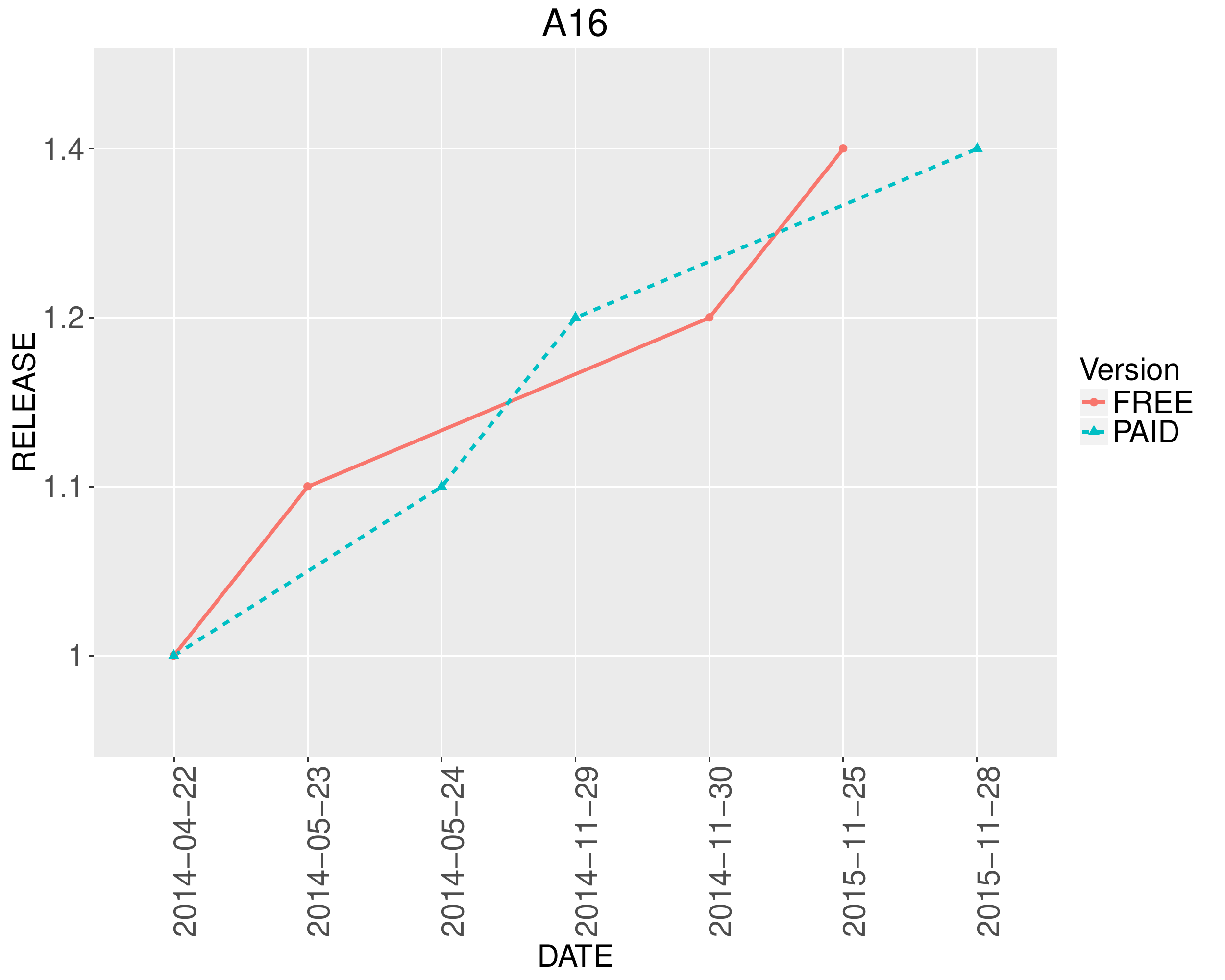}
    \includegraphics[width=0.8\linewidth]{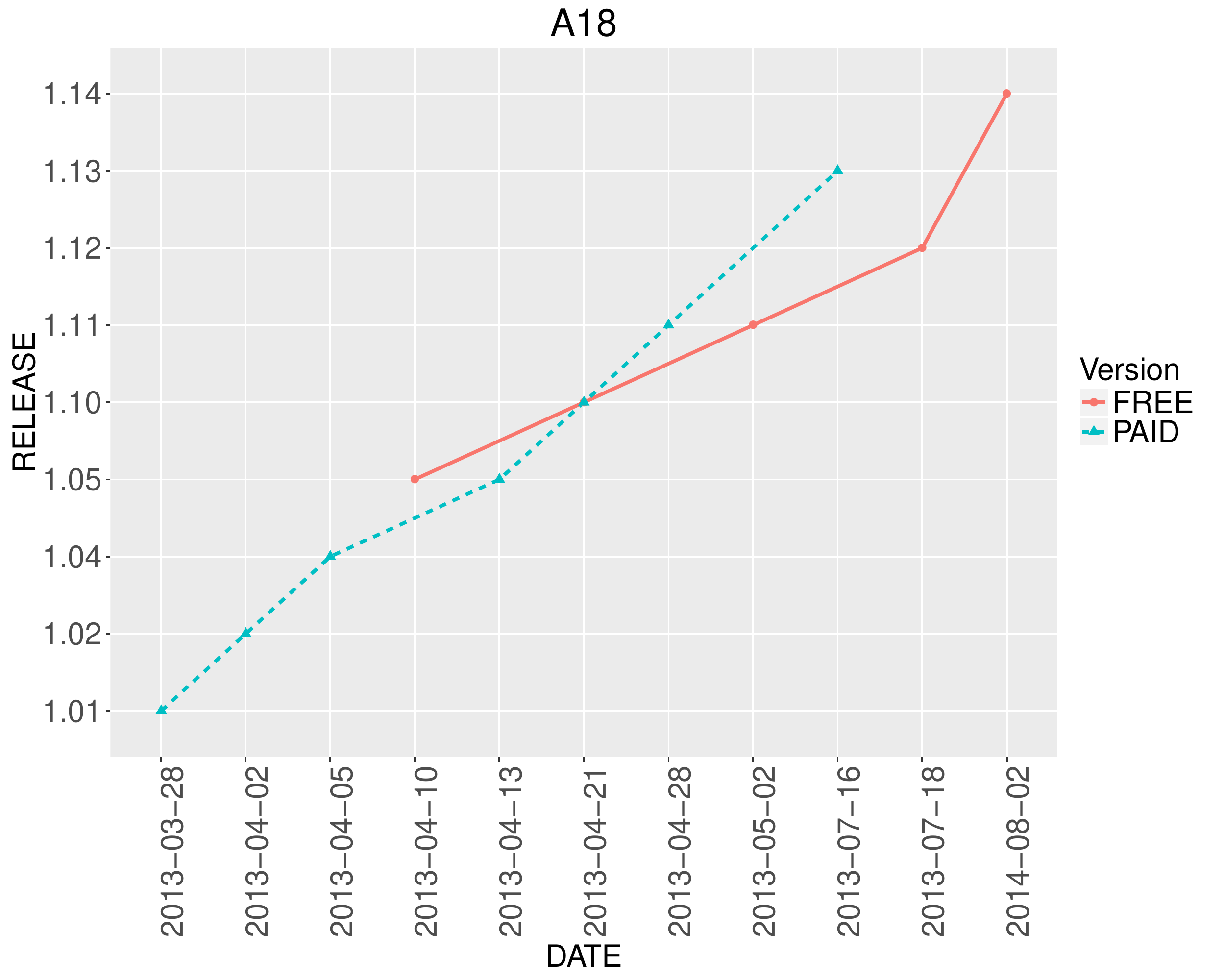} 
  \caption{Comparison of releasing frequencies for three apps.}
  \label{fig:releaseFrequencyForSomeApps} 
\end{figure}

Fig.\ \ref{fig:releaseFrequencyForSomeApps} shows the comparison of releasing frequencies for three apps, for space constraints, chosen among the apps selected for this study. We chose these three apps because they illustrate different ways of releasing apps. For each app is shown the date of each release and the release number, for both ads-supported and paid apps. For A05, the ads-supported version was released before the paid one. Then, when the free app was mature enough, the paid app was released and, from this point in time on, both ads-supported and paid apps were released concurrently. When considering the A16, the first version for both ads-supported and paid versions were released at the same time. Then, ads-supported and paid apps were released alternatively with a difference of few days between them. For A18, the paid app was released and, one month later, the free version was released, probably due to a small number of downloads for the paid version.

\hypobox{Ads-supported apps usually have more releases and are released more often than their corresponding paid versions.} 

\subsection*{RQ 4: \rqfour}

\subsubsection*{Motivation}

Paid apps are not free by definition and have a price that can be changed by developers at any time. We study the evolution of the prices of different releases of paid apps to understand how developers set and update prices.

\subsubsection*{Approach}

The information about the releases of apps as explained above also contained the price of paid apps for each of their releases. Using this information, we can study the evolution of prices of paid apps.

\subsubsection*{Results}

\begin{figure}[!htb] %!htb 
    \centering
    \includegraphics[scale=0.35]{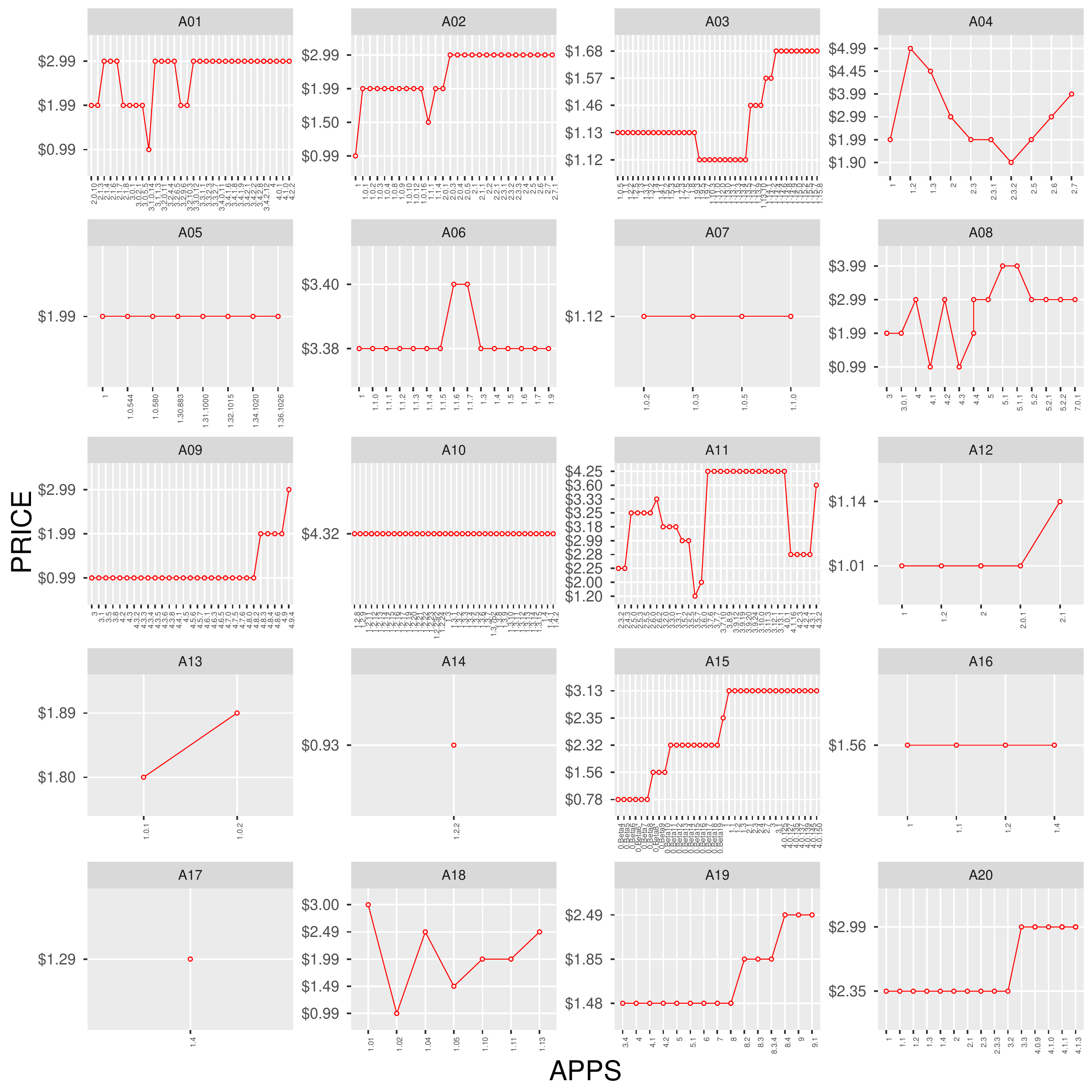} 
    \caption{Price evolution of paid apps.} 
    \label{fig:priceEvolution} 
\end{figure}

Fig. \ref{fig:priceEvolution} shows the evolution of the cost for each paid app and release. It does not show a clear trend. A05, A07, A10, and A16 maintained their prices constant over different releases while prices were increased for A09, A15, A19, and A20. For the other apps the price fluctuated over different releases.

\hypobox{There is no a clear trend in the evolution of prices of paid apps over different releases.}

\subsection*{RQ 5: \rqfive}

\subsubsection*{Motivation}

Developers introduce ads in their free apps to get a revenue. They release paid versions without ads and, sometimes, with more features. We compare the features of ads-supported and paid apps to know if ads-supported apps have as much features as the paid ones.

\subsubsection*{Approach}

Information about the existence of ads or feature limitations in ads-supported apps is available in the descriptions of the apps in Google Play. For each app, we manually check the app description to assess the features offered by both the ads-supported and paid versions of the apps.

\subsubsection*{Results}

In six cases (A03, A10, A14, A15, A18, and A20), paid apps offer more features than their ads-supported versions. It means that only 30\% of the studied apps offer more features in their paid versions. In the majority of cases (70\%), the ads-supported and paid versions of an app are identical in terms of features. Therefore paid apps never offer less features than their corresponding ads-supported versions, which is expected.

\hypobox{Paid apps do not usually include more features than their corresponding ads-supported versions.}

\subsection*{RQ 6: \rqsix}

\subsubsection*{Motivation}

Ads-supported apps use ad networks, which allow developers to include ads in their apps by providing API and content. Some ad networks, such as AdMod, offer ads mediation\footnote{https://firebase.google.com/docs/admob/android/mediation}, which lets developers serve ads to their apps from multiple networks increasing the monetization of their apps by sending ad requests to the ad network with the highest bid. 
When using ad networks, developers set a refresh rate defining how often ads are reloaded. The lower the value the more often are ads reloaded. 
We investigate the numbers of ad networks and the values of the refresh rates used by developers in their apps. 

\subsubsection*{Approach}

We analyze the numbers of ad networks used by apps using Appbrain Ad Detector, which reports over 70 different aspects of apps, including which ad networks are embedded in apps. For each app, for both their ads-supported and paid versions, we retrieve the ad networks that they use. Also, we run each ads-supported app for three minutes and we measure the time between each ad reload to determine their refresh rates.

\subsubsection*{Results}

The number of ad networks used by ads-supported apps is in the range $[1,6]$ (average 1.80). The ad network AdMob, which is powered by Google, is used in all the ads-supported apps with 11 of them (55\%) using this ad network exclusively.

\hypobox{The ad network provided by Google is the most used in ads-supported apps, although developers sometimes include additional ad networks.}

Considering paid apps, among the 20 apps used in our study, six paid apps (30\%) contain ad networks and these networks are identical to those in their corresponding free versions. Specifically, A01, A05, A08, A15, A16, and A19, include ad networks in their paid versions, although the official Android documentation\footnote{https://firebase.google.com/docs/admob/android/lite-sdk} states that the usage of ad networks increases the sizes of apps and users often avoid downloading large apps\footnote{https://developer.android.com/topic/performance/reduce-apk-size.html}. 

\hypobox{Developers do not always remove ad networks in paid versions of their ads-supported apps, which increases the apps sizes.}

Regarding the refresh rate of ads, ad networks recommend to set it to a value in the range of $[30, 120]$ seconds, although the default value use to be 60 seconds. A zero value means that ads would only be loaded once. We observe that only three apps (15\%) have null refresh rates. For nine apps (53\%), developers set the refresh rate to 60 seconds while they set 30 seconds for seven apps (41\%). Only one app (6\%) has a different refresh rate, in this case of 40 seconds. Thus, most developers keep the default refresh rate proposed by the ad network\footnote{https://support.google.com/admob/answer/3245199?hl=en} while few developers use the minimum refresh rate proposed by ad networks.

We estimate the average refresh rate of ads as $\lceil\frac{100}{1.88}\rceil=56$ seconds. Where 1.88 is the average number of ads reloaded in 100 seconds by ads-supported apps. This average refresh rate value has been estimated omitting the three apps with a null value.

\hypobox{Most developers use the default refresh rate for ads-supported apps.}

\subsection*{RQ 7: \rqseven}

\subsubsection*{Motivation}

Permissions is a mechanism that enforces restrictions on the operations that apps can perform. They are requested in the manifest file of the app. 
Book \etal~\cite{book_longitudinal_2013} showed that ad networks are increasingly taking advantage of the permissions requested by the app, possibly to compromise the users' privacy. 
As it is said in the Android documentation\footnote{https://goo.gl/jckp86}, it is strongly recommended to minimize the number of permissions that apps request. We compare ads-supported and paid apps in terms of numbers and types of granted permissions.

\subsubsection*{Approach}

Requested and granted permissions can be obtained using the Android tools \texttt{aapt} and \texttt{dumpsys}, respectively. We developed a Python script that obtains, from the \texttt{apk} file of each app, the list of granted permissions for both ads-supported and paid apps. Then, we apply the \texttt{diff} command to analyze the differences between ads-supported and paid versions of each app. Finally, we compute, for the paid apps, the numbers of granted permissions removed and the numbers of granted permissions added in comparison to their corresponding ads-supported versions.

\subsubsection*{Results}

Table \ref{table:permissionAddedPaidApps} shows that, for three apps (15\%), the paid versions include new permissions that do not exist in the ads-supported versions. CHECK\_LICENSE is used to apply license controls to apps published through Google Play. 
GET\_ACCOUNT, MANAGE\_ACCOUNTS, and USE\_CREDENTIALS are permissions related to the Android \texttt{AccountManager} class for logging in with Google and validate the user. 
The READ\_PHONE\_STATE permission allows read only access to phone state and it is used in combination with CHECK\_LICENSE permission for licensing validation.

\begin{table}[h]
\centering
\caption{Apps that grant more permissions for the paid version.}
\label{table:permissionAddedPaidApps}
\begin{tabular}{clc}
\hline
App & Permissions added to the paid version                                                                                                                                 \\
\hline
A02 & com.android.vending.CHECK\_LICENSE                                                                                                                                       \\
\hdashline[1pt/1pt]
A03 & \begin{tabular}[c]{@{}l@{}}android.permission.GET\_ACCOUNTS\\ android.permission.MANAGE\_ACCOUNTS\\ android.permission.USE\_CREDENTIALS\end{tabular} 
                     \\
                     \hdashline[1pt/1pt]
A10 & \begin{tabular}[c]{@{}l@{}}android.permission.READ\_PHONE\_STATE\\ 
com.android.vending.CHECK\_LICENSE\\ %com.android.launcher.permission.INSTALL\_SHORTCUT
\end{tabular}     \\
\hline
\end{tabular}
\end{table}

For seven apps (35\%), both ads-supported and paid versions have exactly the same numbers and types of permissions. Concerning paid versions, for 10 apps (50\%), permissions are removed in comparison to ads-supported versions (mostly permissions related to network access and location, which are required to load ads). Specifically, for A04, A09, A11, A14, A17, and A18, the paid versions require less permissions than the ads-supported ones because the former do not need to access to the Internet, but it is required when ads are included in the free version. Table \ref{table:permissionRemovedPaidApps} shows the list of paid apps that grant less permissions than their corresponding ads-supported versions.

\begin{table}[t]
\centering
\caption{Apps that grant less permissions for the paid version.}
\label{table:permissionRemovedPaidApps}
\begin{tabular}{cl}
\hline
App & Permissions removed in the paid version                                                                                                                                                                                                                                      \\
\hline
A01   & \begin{tabular}[c]{@{}l@{}}com.google.android.providers.gsf.permission.READ\_GSERVICES\\ android.permission.READ\_PHONE\_STATE\end{tabular}                                                                                                                                                                               \\
\hdashline[1pt/1pt]
A04   & \begin{tabular}[c]{@{}l@{}}android.permission.INTERNET\\ android.permission.ACCESS\_NETWORK\_STATE\end{tabular}                                                                                                                                                                                                              \\
\hdashline[1pt/1pt]
A06   & \begin{tabular}[c]{@{}l@{}}com.google.android.providers.gsf.permission.READ\_GSERVICES\\ com.android.vending.BILLING\\ android.permission.READ\_PHONE\_STATE\\ com.google.android.gms.permission.ACTIVITY\_RECOGNITION\\ android.permission.ACCESS\_WIFI\_STATE\end{tabular} \\
\hdashline[1pt/1pt]
A09   & \begin{tabular}[c]{@{}l@{}}android.permission.INTERNET\\ android.permission.ACCESS\_NETWORK\_STATE\end{tabular}                                                                                                                                                                                                           \\
\hdashline[1pt/1pt]
A11  & \begin{tabular}[c]{@{}l@{}}android.permission.INTERNET\\ android.permission.ACCESS\_NETWORK\_STATE\end{tabular}                                                                                                                                                                                                             \\
\hdashline[1pt/1pt]
A14  & \begin{tabular}[c]{@{}l@{}}android.permission.INTERNET\\ android.permission.ACCESS\_NETWORK\_STATE\end{tabular}                                                                                                                                                                                                              \\
\hdashline[1pt/1pt]
A16  & android.permission.ACCESS\_COARSE\_LOCATION                                                                                                                                                                                                                                                                                 \\
\hdashline[1pt/1pt]
A17  & \begin{tabular}[c]{@{}l@{}}android.permission.INTERNET\\ android.permission.ACCESS\_NETWORK\_STATE\end{tabular}                                                                                                                                                                                                            \\
\hdashline[1pt/1pt]
A18  & \begin{tabular}[c]{@{}l@{}}android.permission.INTERNET\\ android.permission.ACCESS\_NETWORK\_STATE\end{tabular}                                                                                                                                                                                                              \\
\hdashline[1pt/1pt]
A20  & \begin{tabular}[c]{@{}l@{}}com.android.vending.BILLING\\ android.permission.READ\_EXTERNAL\_STORAGE\\ android.permission.ACCESS\_COARSE\_LOCATION\\ android.permission.WRITE\_EXTERNAL\_STORAGE\end{tabular}                                                                                                               \\
\hline
\end{tabular}
\end{table}

The READ\_PHONE\_STATE and the READ\_GSERVICES permissions are required by some ads providers to get some information about the phone and read Google service configuration data. The BILLING permission is needed for sending in-app billing requests and managing in-app billing transactions using Google Play. Ads-supported versions of A06 and A20 need this permission because they offer the possibility of unlocking additional content. The ACTIVITY\_RECOGNITION permission is required by app A06 to integrate additional ads SDK into the AdMob advertising platform. Permissions to read from and write to external storage are granted by some ads providers (as MoPub or MillennialMedia). Even if they are not mandatory, ads providers recommend them but developers can choose to use them or not.

\hypobox{In general, paid apps require less permissions than their corresponding ads-supported versions. Yet, the validation of their licensing may require extra permissions which are not needed by their corresponding ads-supported versions.}

\section{Experimental Study}
\label{sec:experiments}
We answer our second high-level research question, {\em \rqmainb}, through an experimental study over the ads-supported and paid apps selected and shown in Table \ref{table:applications}. Past studies \cite{wei_profiledroid:_2012,gui_truth_2015} discussed the impact and hidden costs of ads on performance metrics. However, they did not support their reports using statistical tests and, therefore, their conclusions could be statistically invalid. We complement these previous studies with measures that we can analyze statistically to check if ads-supported apps are more costly in terms of performance metrics than paid apps because of the presence of ads. In addition, we propose different equations to determine the cost of free apps due to ads.

For each app, we create a simple scenario to start the app, skip the initial tutorial (if present), and wait for 100 seconds in the main activity that contains ads in the free version. These scenarios are run 30 times automatically while we collect performance metrics. 
Before the runs, the screen brightness is set to the minimum value and the phone is set to keep the screen on. To avoid any kind of interference during the measures, only the essential Android services are run on the phone.

We collect the following performance metrics (CPU, memory, network usages, and power consumption) as follows.

\subsubsection*{CPU usage}

It is collected using the approach proposed by Gui \etal \cite{gui_truth_2015}. The \texttt{top} command is run on the phone in the background, obtaining the percentage of CPU usage associated to an Android app. Every second, this information is added to a file stored on the phone.

\subsubsection*{Memory usage}

It is measured using the \texttt{dumpsys meminfo} command on the phone. Every second, this information is obtained for the process associated with the Android app and added to a file stored on the phone. We measure memory using the Proportional Set Size (PSS), which is proposed in the Android documentation\footnote{https://developer.android.com/studio/profile/investigate-ram.html} and which is a measure of the app's RAM use, including shared pages across processes. 
This metric is different from the one used by Gui \etal \cite{gui_truth_2015}, where the Resident Set Size (RSS) was used, which indicates how many physical pages are associated with a process and which is less precise. 

\subsubsection*{Network usage}

It is collected using the \texttt{tcpdump}\footnote{http://www.androidtcpdump.com/} command on the phone as in recent works \cite{gui_truth_2015,saborido_optimizing_2016}, which captures packets from the WiFi and cellular connections. We use this tool via \texttt{adb} to capture the numbers of bytes transmitted over the network connection.

\subsubsection*{Power consumption}

\begin{figure}[!htb]
\centering
\includegraphics[scale=0.2]{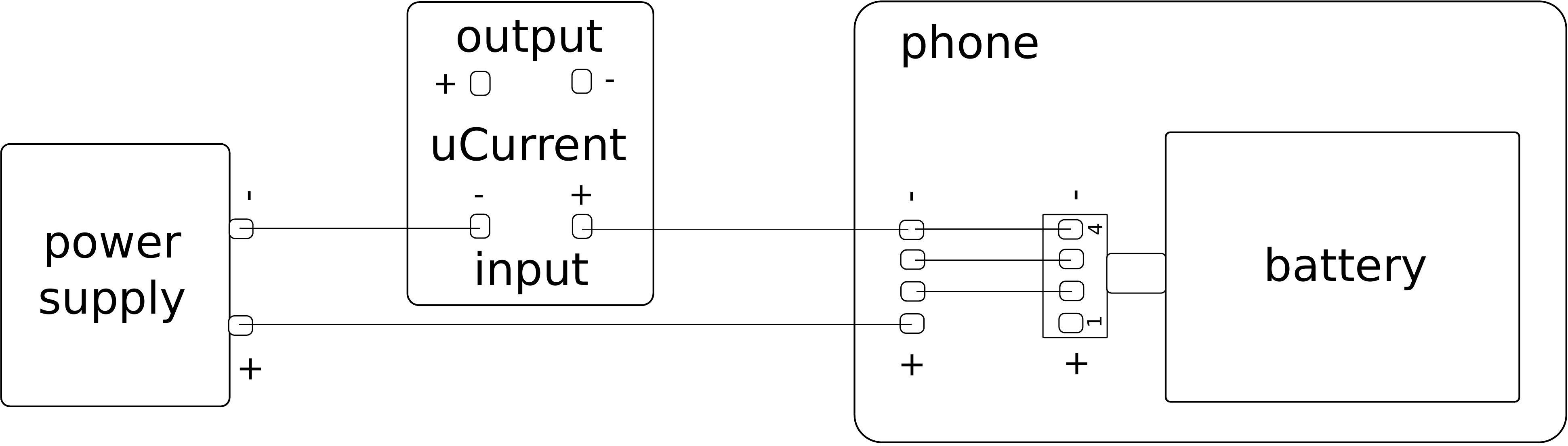}
\caption{Connection between power supply and the Nexus 4.}
\label{fig:circuit}
\end{figure}

It is measured using a digital oscilloscope, a TiePie Handyscope HS5, which offers the LibTiePie SDK, a cross-platform library for using TiePie engineering USB oscilloscopes in third party applications. We use this device because it allows higher frequencies than the Monsoon Power Monitor\footnote{https://www.msoon.com/LabEquipment/PowerMonitor/}. We set the resolution of the oscilloscope to 16 bits and the frequency to $125kHz$. Therefore, the device takes a measure each eight microseconds. 
The mobile phone is powered by a power supply and, between both we connect, in series, a uCurrent\footnote{http://www.eevblog.com/projects/ucurrent/} device, which is a precision current adapter for multimeters, converting the input current $I$ proportionally to the output voltage $V_{out}$. 
Knowing $I$ and the voltage supplied by the power supply $V_{sup}$, we use Ohm's law to calculate the power usage $P$ as $P = V_{sup} \cdot I$. The diagram of the connection is shown in \mbox{Fig. \ref{fig:circuit}}.

Because the phone is connected via USB to the PC to send and receive data, we disabled the USB charging on the phone. We developed a simple Android app to enable or disable USB charging in Nexus 4 phones. This app is free and available for download in Google Play\footnote{https://play.google.com/store/apps/details?id=ruben.nexus4usbcharging}.

\begin{figure}[] %!htb
    \centering
    \includegraphics[width=1\linewidth]{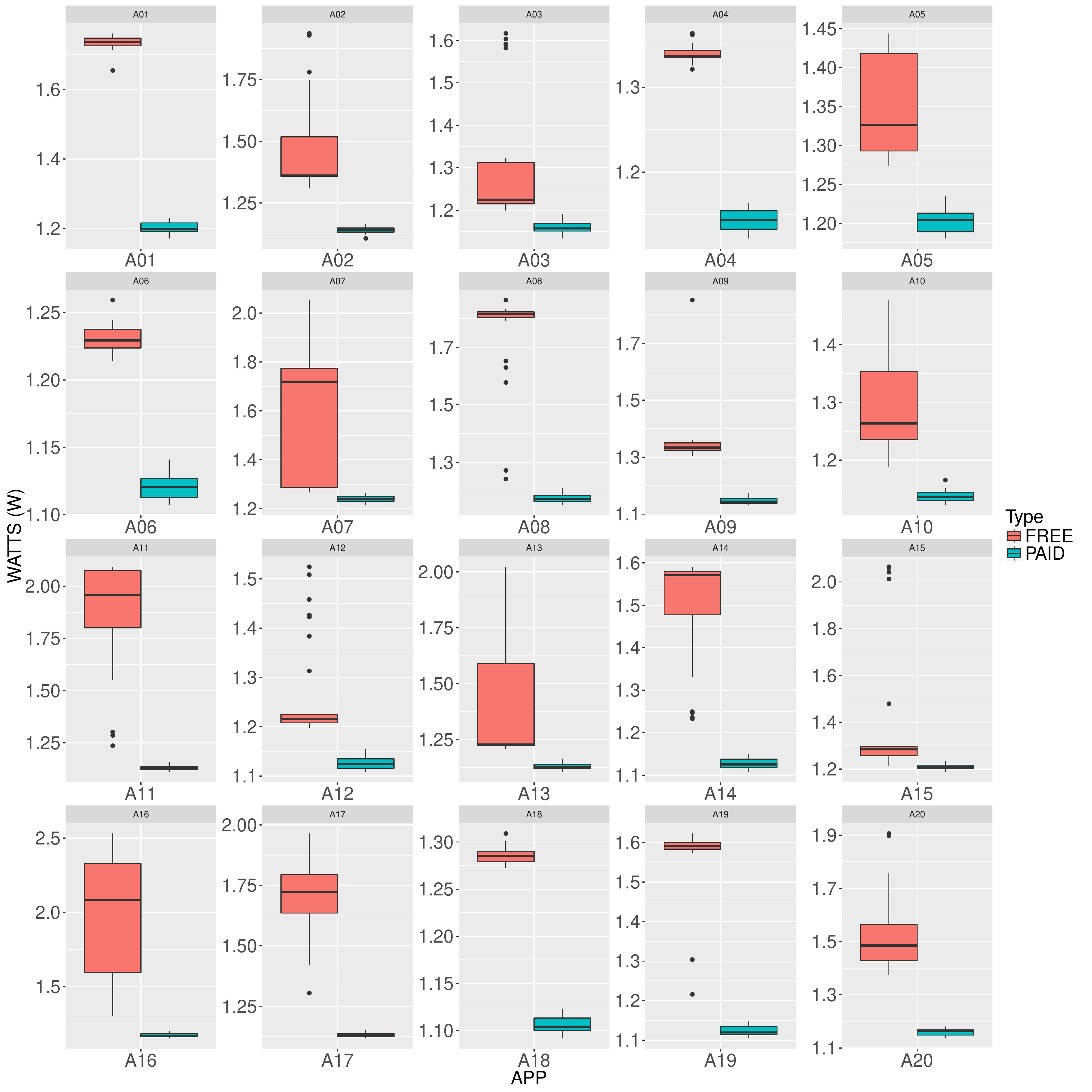} 
    \caption{Power usage by app.} 
    \label{fig:PowerByApp} 
\end{figure}

Fig.\ \ref{fig:PowerByApp} shows the distribution of power consumption for each app individually. We observe that the ads-supported versions are always more power-consuming than the paid ones for the considered measures. A similar trend is obtained for CPU, memory, and network usages (we thus omit the figures).

We perform a Wilcoxon rank sum test to check if the differences observed between the values of the measures of ads-supported and paid versions are significant. Our null hypothesis is that the distributions of the measures of paid apps and that of their corresponding ads-supported versions differ by a location shift of $\mu$ (the average value), expecting that paid apps have a better performance. We consider the difference to be significant if the obtained $p$-value is lower than 0.05. In addition, we compute the effect size using the Cliff's Delta function from the \texttt{R} software \footnote{https://cran.r-project.org/web/packages/effsize/} when the comparison is significant. 

We do not get a significant difference and the null hypothesis cannot be rejected for only one app (A01) and only one metric (memory usage). For all the other apps and measures, there are statistically significant differences with a large effect and we can reject the null hypotheses.
We conclude that paid apps are more efficient in terms of CPU, memory, network, and power consumption because of the absence of ads. 

\begin{figure}[h] %!htb 
    \centering
    \includegraphics[width=0.6\linewidth]{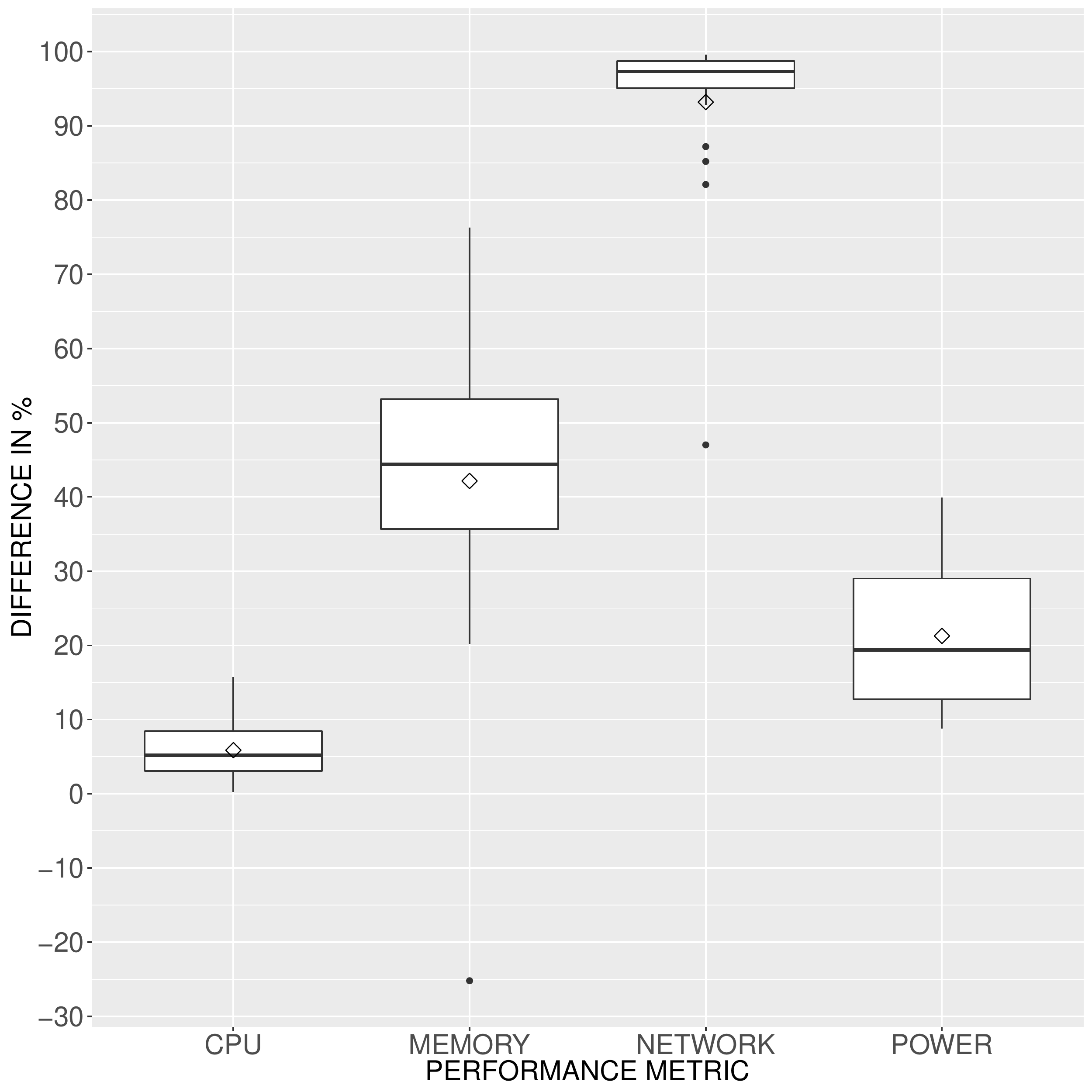} 
    \caption{Performance metric differences for ads-supported and paid apps.} 
    \label{fig:freeVSnonfree} 
\end{figure}

Fig. \ref{fig:freeVSnonfree} shows the differences in percentages for each performance metric between ads-supported and paid apps, where the symbol $\lozenge$ shows the average value. 
Ads-supported apps consume, in average, 5.88\%, 42.15\%, 93.19\%, and 21.27\% more than paid apps for CPU, memory, network, and power consumption, respectively.

\hypobox{Ads-supported apps use more resources than their corresponding paid versions and differences are statistically significant and the effect size large.}

From the collected data, we obtain that, in average, ads-supported apps increase network usage by 1.28MB (median 0.56MB). The average data network usage is so high because A01 consumes a lot of data in its free version (average 14.43MB in comparison to the 0.06MB consumed by the paid app). Using the median value, we estimate the average network usage of each ad. Taking into account that in 100 seconds, in average, 2.6 ads are loaded, we conclude that 0.21MB $\left(\frac{0.56MB}{2.6}\right)$ is the average network usage of each independent ad. We thus estimate the  network cost of ads using the price of \$15 per gigabyte as provided by the AT\&T company, a popular American Internet services provider. We thus determine that each load of an ad potentially costs end users \$0.00315 in network charges. Concerning power consumption, in average, ads-supported apps increases it by 0.34W (median of 0.27W). Instead of power, we estimate the cost of ads in terms of energy consumption. The average difference in terms of energy consumption between ads-supported and paid apps is 36.35J (median 20.29J). Considering this fact  
we conclude that 13.98J $\left( \frac{36.35J}{2.6}\right)$ is the average energy consumption of each independent ad. %\yann{AS with AT company, could we put \$ on the cost of electricity?}

\hypobox{The average network usage and energy consumption of independent ads is 0.21MB and 13.98J, respectively.}

Taking into account the average energy consumption of ads we use Eq. \ref{eq:charge} to calculate the percentage of battery charge that is consumed by ads. $E$ is the average energy consumption of independent ads in joules (J), and $V$ and $C$ are the voltage and electric charge (in mAh), respectively, of the phone battery. Replacing constants by the Nexus 4 values ($V=3.8$ and $C=2100mAh$), we conclude that, in average, each independent ad consumes 0.0486\% of the total battery. 

\begin{equation}
\label{eq:charge}
\%_{charge} = \frac{E}{V} \times \frac{1000}{C \times 3600} \times 100
\end{equation}

We analyze the impact of the refresh rate of ads on the real cost of ads-supported apps calculating the total network (in MB) consumed by ads in ads-supported apps using Eq.\ \ref{eq:network}, where $D$ is the running time of the app (in seconds), $R_{rate}$ is the refresh rate of ads (in seconds), and $network_{ads}$ is the average network usage of ads in MB.

\begin{equation}
\label{eq:network}
network = \left(\frac{D}{R_{rate}}\right) \times network_{ads}
\end{equation}

We define a similar equation to analyze the impact of the refresh rate of ads on battery life for ads-supported apps using Eq.\ \ref{eq:batterydrained}, where $battery_{ads}$ is the average percentage of battery consumed by ads, which can be calculated using Eq.\ \ref{eq:charge}.

\begin{equation}
\label{eq:batterydrained}
battery_{drained} = \left(\frac{D}{R_{rate}}\right) \times battery_{ads}
\end{equation}

In addition, given a data plan size, we calculate the time (in seconds) in which the data limit would be reached because of ads using Eq.\ \ref{eq:dataplan}, where $datasize$ is the size of the data plan in MB, $R_{rate}$ is the refresh rate of ads (in seconds), and $network_{ads}$ is the average network usage of independent ads in MB.

\begin{equation}
\label{eq:dataplan}
D_{dataplan}=\frac{datasize \times R_{rate}}{network_{ads}}
\end{equation}

There is a point in time in which the hidden costs of ads-supported apps overtake the clear costs of paid apps and, thus, ads-supported apps could be more expansive. To estimate the time in which an ads-supported app overtakes its paid version, we define Eq.\ \ref{eq:amortization}, where $price_{app}$ is the cost of the paid app, $R_{rate}$ is the refresh rate of ads (in seconds) in the free app, \mbox{$network_{ads}$} is the average network usage of each independent ad in MB, $price_{MB}$ is the MB price of overage data, and $members$ is the number of people in the Google Family Group sharing the app.

\begin{equation}
\label{eq:amortization}
amortization = \frac{price_{app} \times R_{rate}}{network_{ads} \times price_{MB} \times members}
\end{equation}

Although the cost of network usage and battery life could be considered small, it depends on the use of the app as follows. Let us suppose that we use the ads-supported version of A11, which is a popular PDF book reader. A11 uses a refresh rate of 60 seconds. If a user uses A11 100 minutes every day for a month, Eq.\ \ref{eq:network} shows that about 630MB of data would be spent only in ads. If the data plan is limited, for instance, to 500MB, Eq.\ \ref{eq:dataplan} shows that it would be over in 24 days and the user would be billed for 130MB that supposes \$1.95. Regarding battery life and considering Eq.\ \ref{eq:batterydrained}, because the app is used every day for 100 minutes, the battery percentage would be decreased an additional 4.86\% each day, for only loading ads. Using Eq.\ \ref{eq:amortization}, and considering that A11 costs \$3.49, we obtain that if the data plan is over and the free app is used only by one member of the family group for 19 hours, the cost of data overage due to ads would be higher than the price of the paid app. If we consider that up to five users can belong to a family group, the cost of data overage would be higher than the price of the paid version in only four hours.

\hypobox{Depending on use and service provider, paid apps are sometimes less expensive because their costs are amortized in a short period of time.}

\section{Threats to Validity}
\label{sec:threat}
Threats to internal validity concern factors, internal to our study, that could have influenced our results. We collected information about releases from Google Play and AppBrain. The latter is a public source for information about Android apps and we cannot guarantee that it does not introduce any bias. However, we compared information existing in AppBrain and Google Play for the last version of the apps used in our study and confirmed that the information offered by AppBrain was consistent. We only considered the last versions because they are the only version available in Google Play. To obtain information about the ad networks used in the apps we used AppBrain Ad Detector, which is available for free in Google Play. We verified the information offered by this app using the free app Addons Detector, also available in Google Play. Considering performance metrics, we computed the power and energy consumption using well-known theory and scenarios. In addition, we replicated several times our measures to ensure statistical validity. Concerning the cost of ads-supported apps, given that we cannot accurately simulate all users contexts (user's location and timezone) and because we do not know for how long ads are cached before they are reloaded, we chose to uninstall apps between different runs. Although this fact reduced the memory overload in the real phone during the experiments, it probably artificially increased the network usage because ads were not always cached.

Threats to construct validity concern relationship between theory and observation and the extent to which the measures represent real values. We used a Nexus 4 phone, the same model used in previous studies~\cite{linares-vasquez_mining_2014,sahin_how_2014,huang_defdroid:_2016,saborido_optimizing_2016,sahin_benchmarks_2016}. Our measurement apparatus had a higher or the same number of sampling bits as previous studies and our sampling frequency was one order of magnitude higher than past studies. Overall, our measures were more precise or at least as precise as those in previous studies. Network and CPU usages were collected using the tool \texttt{tcpdump} and the command \texttt{top} on the phone, respectively, which may have introduced extra energy consumption \cite{gui_truth_2015}. We confirmed that running \texttt{tcpdump} and \texttt{top} did not have noticeable impact on energy consumption and we extended this observation to \texttt{dumpsys meminfo}, the command used to collect information related to memory usage.

Threats to conclusion validity concern the relationship between experimentation and outcome. While the analyses related to performance metrics was supported by appropriate statistical procedures, the exploratory study was qualitative in nature and, hence, we did not use any statistical analysis. Our estimations of the hidden cost of ads and, therefore, of the cost of ads-supported apps were based on the data collected in the experimental study, which was limited to 20 apps and to the Nexus 4 phone. Further validations on a larger set of apps and--or different phone is desirable to make our findings more generic. 

Threats to external validity concern the generalization of our findings. The study was limited to a reduced number of apps. We consequently must accept the App Sampling Problem \cite{martin_app_2015}, which exists when only a subset of apps are studied, resulting in potential sampling bias. We selected apps belonging to different categories and developed by different developers.  
More experiments over larger set of apps must be carried out to generalize our findings to the entire Android ecosystem.

We ensure the replicability of our study by making all collected data and scripts available in a replication package\footnote{It will be available in June 2017.}. In addition, all the figures and tables omitted in the paper, because of the space limitation, are also available in our replication package.

\section{Related Work}
\label{sec:related-work}
Power consumption and data traffic of mobile apps are nowadays a hot topic given the popular use of mobile devices and these issues have been addressed in many papers \cite{linares-vasquez_optimizing_2015,hindle_greenminer:_2014,pathak_where_2012,saborido_optimizing_2016}, just to name a few contributions. In addition to power consumption and network usage, other metrics have been analyzed in Android apps. In \cite{wei_profiledroid:_2012} was presented ProfileDroid, a monitoring and profiling system for characterizing Android app behaviors at multiple layers: static, user, OS and network. In addition, the impact of ads on performance has been studied in some recent works. In \cite{wei_profiledroid:_2012} is observed that ads-supported versions of apps could end up costing more than their paid versions especially on limited data plans, due to increased advertising/analytics traffic. In \cite{gui_truth_2015} is studied 21 Android apps and they shown that the use of ads leads to mobile apps that consume significantly more network data and have increased energy consumption. They also found that these hidden costs can impact the ratings given to an app. 

Permissions and its relation with ad networks has been also studied. In \cite{book_longitudinal_2013} is shown that the use of most permissions has increased over the last several years, and that more libraries are able to use permissions that pose particular risks to user privacy and security.

There are several approaches focusing narrowly on how to improve mobile ads. In \cite{khan_mitigating_2012} a prototype was implemented with prefetching and caching techniques to improve the energy consumption and network usage of ads. In \cite{vallina-rodriguez_breaking_2012} an ad delivery framework was proposed, which predicts user context to identify relevant ads. More recently, in \cite{pamboris_ad-apt:_2016}, was proposed AD-APT, a system to avoid the adverse implications of mobile advertising on device energy and network usage. AD-APT strikes a balance between these two performance metrics refactoring ad-supported apps automatically to adjust the frequency of mobile ad occurrences at runtime. 

To the best of our knowledge, the ads-business model and differences between ads-supported and paid Android apps have not been explored before. Although it is obvious the impact of ads on performance metrics it was not studied if this impact is significant or not. We do that and, in addition, we propose few equations to get an estimation about the cost of ads-supported apps.

\section{Conclusion}
\label{sec:conclusion}
The Android market is a place where developers offer paid and--or free apps to users. Free apps can follow the freemium or the ads-business model . While the latter includes ads to allow developers to get a revenue, the former offers less features and the user is charged for unlocking additional features.

In this paper, we focused on the ads-business model to understand the differences in the implementation and development process of ads-supported and paid apps. We asked ``\rqmain''. To answer our question, we asked ``\rqmaina'' and ``\rqmainb''. 

For the first question, we conducted an exploratory study on a set of 40 apps from Google Play. We bought 20 paid apps and we downloaded their corresponding ads-supported versions. 
We observed that (i) although paid apps have a better rating ads-supported apps are preferred by users, (ii) developers do not usually offer a paid app without a corresponding free version, (iii) paid apps usually have less releases and are released less often than their corresponding free versions, (iv) there is no a clear strategy about prices set by developers to paid apps, (v) ads-supported apps do not usually include less features than their corresponding paid versions, (vi) developers do not always remove ad networks in paid versions of their ads-supported apps, and (vii) ads-supported apps usually require more permissions than paid apps. 

To answer the second question, we carried out an experimental study showing that ads-supported apps use more resources than their corresponding paid versions with statistically significant differences. We estimated that the average network usage and energy consumption of ads is 0.21MB and 13.98J, respectively. Finally, based on that, we offered different equations to estimate (i) the network usage of ads-supported apps, (ii) the percentage of battery drained due to ads in free apps, (iii) the time in which a data plan is over due to the presence of ads in free apps, and (iv) the time in which an ads-supported app overtakes its paid version. 

We conclude that ads-supported and paid apps are not so different because developers usually start releasing free apps and later modify these apps to release them into the marketplace as paid versions. From our observations we advise developers to: (1) remove ad networks in paid apps because they increase the app size and the number and types of Android permissions, (2) take into account the number and type of required permissions when a license validation approach is used, (3) consider the impact on ads-supported apps performance due to the refresh rate of ads. 

We recommend developers to use approaches as AD-APT, which offers a choice between basic policies to adjust the frequency of mobile ad occurrences. A faster way of reducing the cost of ads-supported apps is setting the refresh rate of ad networks to higher values. Although this action makes ads-supported apps more efficient, it could have an impact on the chance of getting revenues. For this reason more studies about the trade-off between performance of ads-supported apps and revenues due to ads should be carried out as future works. In addition to this, we want to research most popular ad networks and their impact on permissions, performance, and revenues. We also plan to study and compare Android licensing validation approaches, usually used by paid apps, in terms of permissions and performance.

\newpage
\balance
\bibliographystyle{IEEEtran}
\def\IEEEbibitemsep{0pt}
\bibliography{mobile,rating} 

\end{document}